# Coulomb coupling between two nanospheres trapped in a bichromatic optical tweezer

Q. Deplano,[1,3] A. Pontin,[4] A. Ranfagni,[1] F. Marino,[3,4] F. Marin,[1,2,3,4,*]

[1] *Dipartimento di Fisica e Astronomia, Università di Firenze, via Sansone 1, I-50019 Sesto Fiorentino (FI), Italy*
[2] *European Laboratory for Non-Linear Spectroscopy (LENS), via Carrara 1, I-50019 Sesto Fiorentino (FI), Italy*
[3] *INFN, Sezione di Firenze, via Sansone 1, I-50019 Sesto Fiorentino (FI), Italy*
[4] *CNR-INO, largo Enrico Fermi 6, I-50125 Firenze, Italy*
*francesco.marin@unifi.it*

**Abstract:** Levitated optomechanics is entering the multiparticle regime, paving the way for the use of arrays of strongly coupled massive oscillators to explore complex interacting quantum systems. Here, we demonstrate the trapping of two nanospheres inside a dual optical tweezer generated by two copropagating lasers operating at different wavelengths (1064 nm and 976 nm). Due to the chromatic aberration of the tweezer optics, two focal points are created approximately 9 $\mu$m apart, each one acting as an optical trap for a silica nanoparticle. At this distance, the surface charges on the nanospheres produce a Coulomb force that couples their motion along the tweezer axis. The strong coupling regime is achieved, as evidenced by the observed avoided crossing of the normal-mode frequencies. These results highlight the potential of our experimental scheme for future studies on systems of strongly coupled oscillators, including their implementation in optical cavities, both in the classical and in quantum regime.

## 1. Introduction

The harmonic oscillator model is the basis of statistical mechanics, molecular modeling, acoustics and many further topics describing oscillating behaviors. On the other hand, in quantum mechanics the theoretical and experimental exploration of the 2-level system characteristics led to enormous advances in our understanding of atomic and molecular quantum optics, spin 1/2 physics and related mathematical models. All in all, both basic systems have driven the progress in an extremely rich and fertile direction.

It is also interesting to notice how deep analogies between quantum and classical toy-systems can be. In a very pedagogical way, L. Novotny [1] discussed the similarities between a quantum two-level system and a pair of coherently coupled classical harmonic oscillators (HO). He derived analytical expressions for the transfer probability between the two normal modes of the HOs pair during diabatic or adiabatic passage through their resonance, and retrieved similar expressions as in the Laudau-Zener theory for quantum systems [2–6]. Earlier authors already used the later theory to derive the same kind of expressions for two pendulums weakly coupled by a spring [7]. During the 2010's, other authors [8] experimentally worked with two coupled flexural modes of a nanobeam brought to cryogenic temperature, and with this classical two-level system performed Rabi oscillations, Ramsey fringes and Hahn echo experiments [9] as well as classical Stückelberg interferometry [10]. The works cited so far enlighten very well how some quantum features could be extended to classical systems. In recent years, other kind of mechanical oscillators, namely individual dielectric nanoparticles levitated by optical tweezers in vacuum [11], has started to be explored, and are drawing more and more attention in the physics community. Indeed, they essentially behave as point-like 3-dimensional harmonic oscillators, free of mechanical bounds. Moreover, their spring constant can be tuned by simply changing the tweezer optical power. This simplicity and tunability make them ideal candidates for fundamental physics experiments, both in the classical [12–15] and in the quantum regime [16–18]. In the very last years, the experimental realization of multiparticle trapping, with coherent interparticle interaction, has been reported [19–22]. In reference [19], the authors demonstrated optical binding between the center of mass (CoM) of two particles, for which the non-reciprocal component of the interaction paves the way to the exploration of non-Hermitian physics on such elementary systems [23, 24]. Moreover, the optical coupling can be turned on and off by changing the dipoles orientation (i.e., the tweezers polarization). By turning off the optical binding, they could also highlight the Coulomb coupling due to the electrical charges on the particles surfaces, as schematically shown in Fig. 1(a). This is the type of coupling we are discussing in this article. While the authors of [19] realized a pair of parallel tweezers using a spatial light modulator, as previously done in cold atoms experiments [25, 26], we present here an original and simple method to trap several particles along a single axis.

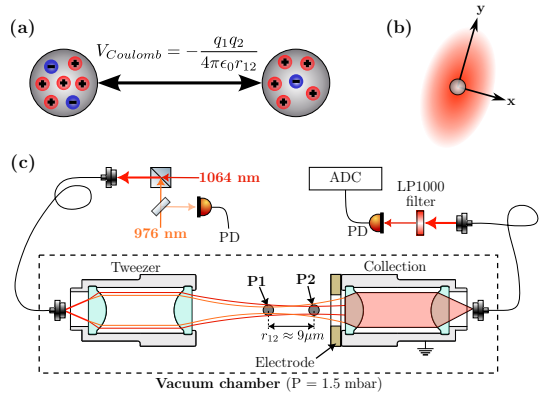

Fig. 1. **(a)** Pair of particles experiencing Coulomb coupling due to their surface charges. **(b)** Transverse cross section of a trapping beam in the focal plane. The elongated shape is mostly due to the projection of the linearly polarized field along the optical axis caused by the high numerical aperture of the focusing lens. **(c)** Experimental setup. The two trapping lasers are mixed in a polarizing beam splitter (PBS) with orthogonal polarizations. A first photodiode (PD) is used to measure the 976 nm laser power. The two laser beams are then brought into a vacuum chamber via a fibered feedthrough terminated by a collimator/focuser which creates the two optical traps. We call P1 the particle trapped in the 976 nm potential and P2 the one trapped in the 1064 nm light. For such wavelength difference, the interparticle distance is about 9 $\mu$m. The tweezer is mounted on a three-axes nanometric positioner allowing for fine alignment with the collection optics. The collected light is filtered with a long pass (LP) filter at 1000 nm and the photo-detection signal is measured with an analog-to-digital converter (ADC).

## 2. Experimental setup

The idea is to use two lasers with sufficiently different wavelengths so that the chromatic aberration of the tweezer focusing optics leads to adequately separated focal points. In this way, two trapping potentials are created along the tweezer axis (Z-axis) at a distance $r_{12}$ that depends on the wavelength difference. The scheme in Fig. 1(c) provides more details about our experimental setup. The light from a semiconductor laser at 976 nm and a Nd:YAG laser at 1064 nm is coupled to an optical fiber terminated by a collimator/focuser consisting of two aspheric lenses with focal lengths of 18.4 mm and 3.1 mm. For our wavelength difference, $r_{12}$ is around 9 µm and one single particle is brought in each trap (see [27] and Supplement 1 for further details about the loading procedure). At this distance the optical binding (dipole-dipole interaction) can be neglected, as discussed in Supplement 2. The traps geometry is schematically shown in Fig. 1(b). The two optical traps are elongated along their respective polarization direction due to the relatively high numerical aperture (0.68) of the tweezer focusing lens [28]. The orientation of the polarization sets the eigenbase for the CoM motion in the transverse plane and the frequencies along the X and Y axis are non-degenerate. We choose X to be the direction of higher CoM frequency for the particle trapped in the 1064 nm laser (P2). Accordingly, it corresponds to the lowest transverse CoM frequency for the particle P1 since the two lasers are mixed in a polarizing beam-splitter with orthogonal polarizations. To measure the motion of the system, the transmitted light is coupled to a further fibered collimator/focuser (collector) and sent on a photodiode after filtering to transmit the 1064 nm laser light only. The Rayleigh scattering of the particles homodynes the transmitted power which therefore fluctuates at the CoM frequencies. If the two optical systems are perfectly aligned on the same optical axis, only the position fluctuations along the tweezer axis are detected. However, by intentionally misaligning the two lens holders, we can find a configuration where all three spatial directions are coupled into the output fiber with enough modulation depth to be detected, as can be appreciated in the spectrum shown in Fig. 2(a).

On the collector head we installed a ring shaped electrode able to generate an electric field along the Z axis (the electric ground is given by the main body of the holder). This electrode is used during the loading of the optical tweezers (see Supplement 1) as well as to coherently drive the Z-motion of the charged particles. This allows to absolutely calibrate the number of charges, as described in detail in Supplement 3 and firstly reported in [29].

## 3. Coulomb coupling

If the particles are charged, the distance-dependent Coulomb interaction couples their motion and leads to the formation of new eigenmodes, respectively symmetric and antisymmetric. The coupling produces a spectral avoided crossing of the system normal modes when tuning the bare oscillator frequencies. The minimum splitting corresponds to the energy exchange rate between the bare oscillators. The strong coupling regime is reached when such energy exchange rate exceeds the damping rate. As a consequence, the avoided crossing can be spectrally resolved.

More formally, we can obtain analytical expressions for the coupling and normal modes frequencies from the study of two HOs simply coupled by a linear spring, described by the Newton's equations.

$$\begin{aligned} m_1 \ddot{r}_{I1} &= -\left(m_1 \Omega_{I1}^2 + k_I\right) r_{I1} + k_I r_{I2} - m_1 \gamma_{I1} \dot{r}_{I1} \\ m_2 \ddot{r}_{I2} &= -\left(m_2 \Omega_{I2}^2 + k_I\right) r_{I2} + k_I r_{I1} - m_2 \gamma_{I2} \dot{r}_{I2} \end{aligned} \quad (1)$$

with $r_{Ii}$ corresponding to the displacement along the axis $I$ around the equilibrium position, with $I = X, Y, Z$. The numerical index $i = 1, 2$ refers to the particle number, $\gamma_{Ii}$ are the damping rates of each oscillator for each direction, $m_i$ are the particles masses.

The $k_I$ are the equivalent coupling spring constants induced by the Coulomb interaction along the three directions. With our setup geometry, they are given by [19]

$$\begin{aligned} k_Z &= 2 \frac{q_1 q_2}{4\pi \epsilon_0 r_{12}^3} \\ k_X &= k_Y = -\frac{q_1 q_2}{4\pi \epsilon_0 r_{12}^3} \end{aligned} \quad (2)$$

($q_i$ are the particle charges, $\epsilon_0$ the vacuum permittivity).

In order to simplify the model and focus on the main physics involved, we assume identical particles (with mass $m$ and damping rate $\gamma$), and we consider the evolution in the frame rotating at the frequency $\Omega' = \sqrt{\Omega_1^2 + k/m}$ by defining the slowly varying variables $\tilde{r}_i = r_i \exp(i\Omega' t)$

(we are omitting for simplicity the index $I$). The evolution equations, neglecting the terms $\ddot{\bar{r}}_i$ and $\gamma \dot{\bar{r}}_i$, can be written as

$$\dot{\bar{r}}_1 = -\frac{\gamma}{2}\bar{r}_1 + ig\bar{r}_2$$
$$\dot{\bar{r}}_2 = \left(-\frac{\gamma}{2} + i\Delta\right)\bar{r}_2 + ig\bar{r}_1 \quad (3)$$

where we have defined the coupling rate

$$g = \frac{k}{2m\Omega'} \quad (4)$$

and $\Delta = \left(\Omega'^2 - \Omega_2^2\right)/2\Omega' - g$. The normal mode frequencies of the coupled oscillators are

$$\Omega_{\mp} = \Omega' - \frac{1}{2}\left(\Delta \pm \sqrt{\Delta^2 + 4g^2}\right). \quad (5)$$

By tuning the frequency of the second oscillator around $\Omega'$, the minimum splitting between the two eigenfrequencies occurs for $\Delta = 0$ and is equal to the energy exchange rate $2g$. We notice that the resonance frequency of the first oscillator shifted by the coupling spring can be written as $\Omega' = g + \sqrt{\Omega_1^2 + g^2}$.

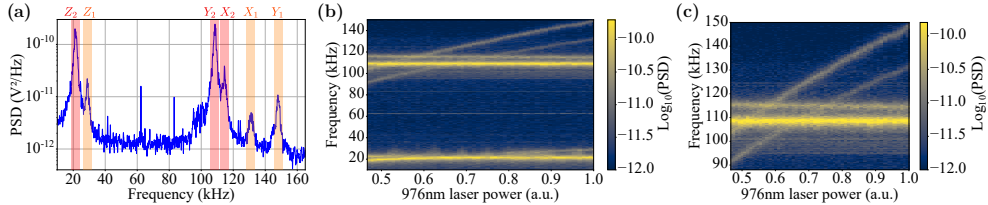

Fig. 2. **(a)** Power spectral density (PSD) of the two particles motion. We can identify six peaks corresponding to the X,Y and Z motion for each particle. **(b)** Evolution of the PSD as a function of the 976 nm laser power. The scans have a duration of 10 seconds and are composed of 60 power steps, each lasting 167 ms. The PSD is computed over the last 150 ms of each step in order to eliminate the power transient between two consecutive steps from the data processing. **(c)** Enlarged view of the crossing between the X and Y modal frequencies of the two particles.

Turning to the experiment, we show in Fig. 2(b) a series of spectra recorded for different values of the 976 nm laser power. We distinguish the pattern corresponding to the six thermally driven CoM motions along the X, Y and Z directions, for each of the two particles. The effect of the power variation on the optical potential felt by particle 2 is negligible, while the CoM frequencies of particle 1 change proportionally to the square root of the laser power. This allows us to sweep them through resonance with the other particle, for all three spatial modes. For the particles used in this experiment, we evaluated $|q_1| = (67 \pm 7)\,q_e$ and $|q_2| = (90 \pm 13)\,q_e$ where $q_e$ is the elementary electrical charge. The two charges have the same sign, yielding repulsive Coulomb force. The particles have a nominal diameter of $125 \pm 5$ nm and by taking an a priori density of 1850 kg/m$^3$ and a distance of $r_{12} = 8.8$ $\mu$m we derive the coupling rates $g_X/2\pi = -114 \pm 24$ Hz, $g_Y/2\pi = -127 \pm 27$ Hz and $g_Z/2\pi = 1270 \pm 270$ Hz. To assess the achievement of the strong coupling regime, these values must be compared to the natural full linewidth of the oscillators. At the working pressure ($p = 1.5$ mbar) they are expected to be $\approx 1.5$ kHz [17, 30, 31]. In these conditions, the strong coupling regime defined by $2g > \gamma$ is reached only for the CoM motion along the Z axis. This is confirmed by the spectra shown in panel (c) of Fig. 2: the X and Y modal frequencies of both traps cross each other. On the contrary for the Z motion, whose spectral signatures are shown in Fig. 3(a), a clear avoided crossing is visible.

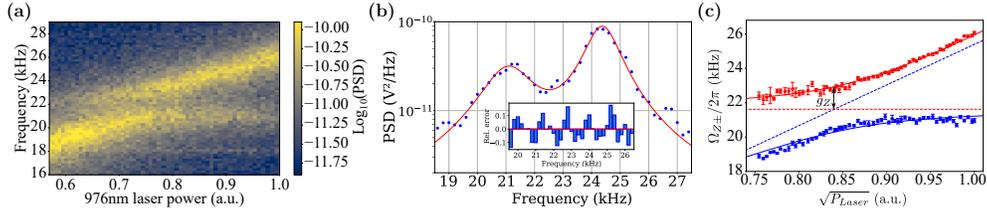

Fig. 3. **(a)** Close-up on the avoided crossing observed between the Z normal modes. This data set was recorded separately from the data shown in Fig. 2(b). Indeed, here we have optimized the scan range and the position between the two lens holders to emphasize this avoided crossing. **(b)** Experimental PSD fitted by a double Lorentzian model used to extract the peak centers (relative residuals are shown in the inset). **(c)** Normal-modes frequencies fitted with Eq. (5) where the coupling rate $g$, the shifted frequency of the first oscillator $\Omega'$, and a proportionality constant between $\Omega_2$ and the square root of the 976 nm laser power are left as free parameters.

To experimentally evaluate the Z coupling rate, we fit the centers of the normal mode spectral peaks using a double Lorentzian model (see Fig. 3(b)). Afterwards the peak central frequencies as a function of the laser power are fitted using equation (5), as shown in Fig. 3(c). Here we have assumed a square root dependence of $\Omega_{Z1}$ on the measured 976 nm laser power. The fitted coupling constant is $g_Z/2\pi = 1323 \pm 38$ Hz, where we are only giving the statistical uncertainty, without considering the systematic error introduced by the approximations that lead to Eq. (5). Using Eqs. (2,4), the measured coupling allows to evaluate the particles separation. Assuming identical masses, we find $r_{12} = 8.7 \pm 0.6\,\mu$m, where the uncertainty is dominated by the knowledge of the charges, in agreement with previous estimates (see Supplement 1).

Although the Lorentzian model fits the spectra with fairly symmetric residuals, as shown in the inset of Fig. 3(b), for strongly coupled oscillators, the expected spectrum is in general not Lorentzian near the resonance. The detected signal can be written as a weighted coherent sum of the individual particles displacement $(Ar_{Z1} + Br_{Z2})$, where $A$ and $B$ are amplitude coefficients depending on the light collection. Returning to the full model, by adding a thermal force noise in equations (1) and after some algebra (see Supplement 4 for further details), we find that the measured spectrum of the coupled system can be expressed as

$$S_{out} = \frac{\left|A\,s_2 + B\,\frac{c}{\mu}\right|^2 + |B\,s_1 + A\,c\mu|^2\,\frac{\gamma_{Z2}}{\gamma_{Z1}\mu^2}}{|s_1 s_2 - c^2|^2} \qquad (6)$$

with $\mu = \sqrt{m_2/m_1}$, $c = k_Z/\sqrt{m_1 m_2}$ and $s_i = -\omega^2 - i\omega\gamma_{Zi} + \Omega_i^2$ for $i = 1, 2$, where we have defined $\Omega_1^2 = \Omega_{Z1}^2 + c\mu$ and $\Omega_2^2 = \Omega_{Z2}^2 + \frac{c}{\mu}$. This expression assumes that the thermal force noise sources driving the motion of the two particles are not correlated.

Fitting the experimental spectra with Eq. (6) allows a direct estimation of the mass ratio and the damping rates acting on the two particles. Assuming $R_1/R_2 = (\gamma_{Z1}/\gamma_{Z2})^{1/2}/\mu$ and $\rho_1/\rho_2 = \mu(\gamma_{Z2}/\gamma_{Z1})^{3/2}$ [30, 31], it becomes possible to differentiate variations of radius $R_i$ and density $\rho_i$, and provide statistical uncertainties of their relative fluctuations (which are usually assumed values) when sampled over different particles.

On the current dataset we extracted a mass ratio $\mu^2 = 1.06 \pm 0.12$ which is thus consistent with 1. On the other hand, the 150 ms time span over which the 976 nm laser power is constant, does not allow a sufficiently accurate estimation of the linewidths ratio. In principle, longer acquisition times would solve the problem and, indeed, uncertainties around 1% were demonstrated [32] by averaging over 60 s at a similar pressure. We could not implement this approach due to slow drifts in the relative alignment between the tweezer and the collector lens, yielding long term fluctuations in *A* and *B*. Of course, better mechanical compliance or a move to free space forward detection [33, 34] will allow to exploit the full potential of the approach.

## 4. Conclusion

In summary, we have demonstrated the Coulomb coupling between the axial CoM motions of two silica nanoparticles trapped in a bichromatic optical tweezer. We evaluated the coupling rate at $1323 \pm 38$ Hz, well above the strong coupling limit. In our original though simple experimental configuration, the two particles are by construction geometrically aligned along the coupled motion axis. As a consequence, there is no need to expand the coupling in Z,X or Z,Y. Moreover, the fibered setup together with the incorporated collimator/focuser guarantees the reproducibility of the laser spatial modes, trap shapes and orientations, and is therefore ideal for robust classical two-level system experiments at room temperature. The oscillators quality factor achieved in our work is of the order of 10, but it can be increased by lowering the pressure in the vacuum chamber. A pressure of $10^{-3}$ mbar is commonly achievable without losing the particles, although, in this first work we kept a relatively high pressure to ensure the robust particle stability even at the lowest power levels of the 976 nm laser. We remark that such experiments can be performed even without highly coherent lasers, since only a relatively high and tunable power is required.

Taking the experiment to the next level, one could detect the motion in free space, perform feedback cooling [21] of the Z motions down to the quantum regime (as it has been already demonstrated on a single levitated particle [35, 36]) and investigate the classical to quantum transition, exploring the conceptual gap between the two paradigms. Such perspective requires low noise lasers and higher detection efficiency than we have here. Additionally, using lasers that are highly tunable, the distance between particles as well as the coupling constant can be tailored, offering even more versatility to the experimental scheme.

An additional interesting benefit of our approach is the possibility to dynamically shift the trap centers by modulating the laser wavelength and thus leveraging the chromatic aberration. By doing it at the Z CoM frequencies, one can selectively drive the motion of the individual particles. Down to the quantum regime, such procedure would be of high importance in order to implement a controlled state preparation protocol able to individually address each nanosphere. A modulation depth of 0.2 pm would produce a displacement of the order of the zero point fluctuations. The length scale of thermal fluctuations is instead achieved with a wavelength variation of 3 nm. Recently, it has been reported the realization of integrated semiconductor lasers self-injection locked to a microcavity, able to scan ∼ 0.1 nm in single mode operation, and 12 nm of course tuning with a bandwidth of 30 kHz [37]. Used together with a tapered fiber amplifier to increase the power, it would be enough to drive the Z degree of freedom with sufficient dynamic range. Simultaneously, one could use an engineered focusing lens that magnifies chromatic aberration, using a high dispersion glass, and increase the effect by up to a factor of three.

Finally, cooling several particles can also be achieved by bringing them in a high finesse optical cavity and exploiting coherent scattering [16–18, 38–43]. The two particles can therefore experience an effective long-range coherent coupling via the cavity mode as is has been realized in [44].

**Funding.**

**Acknowledgments.** We acknowledge financial support from PNRR MUR Project No. PE0000023-NQSTI and by the European Commission-EU under the Infrastructure I-PHOQS "Integrated Infrastructure Initiative in Photonic and Quantum Sciences " [IR0000016, ID D2B8D520, CUP D2B8D520].

**Disclosures.** The authors declare no conflicts of interest.

**Data availability.** Data underlying the results presented in this paper are not publicly available at this time but may be obtained from the authors upon reasonable request.

**Supplemental document.** See Supplemental document for supporting content.

# Coulomb coupling between two nanospheres trapped in a bichromatic optical tweezer: Supplemental document

Q. Deplano,[1,3] A. Pontin,[4] A. Ranfagni,[1] F. Marino,[3,4] F. Marin,[1,2,3,4,*]

[1] *Dipartimento di Fisica e Astronomia, Università di Firenze, via Sansone 1, I-50019 Sesto Fiorentino (FI), Italy*
[2] *European Laboratory for Non-Linear Spectroscopy (LENS), via Carrara 1, I-50019 Sesto Fiorentino (FI), Italy*
[3] *INFN, Sezione di Firenze, via Sansone 1, I-50019 Sesto Fiorentino (FI), Italy*
[4] *CNR-INO, largo Enrico Fermi 6, I-50125 Firenze, Italy*
[*]*francesco.marin@unifi.it*

**Supplementary 1 : Loading of the dual optical tweezer**

To load the dual tweezer with two particles, we first trap one particle on a loading tweezer formed by sending light from an auxiliary 976 nm laser source through the optical system that is later used as light collector during the main experiment. The loading is performed inside a first vacuum chamber, separated from the science chamber by a gate valve which is kept closed during the loading procedure. We then open the gate, move the loading tweezer to the main tweezer optics using a movable arm [1], and superimpose their focal points by maximizing the 976 nm power transmitted through the main tweezer. We then move the main tweezer optics by few microns in order bring its focal point for 1064 nm light close to the particle. The transfer of the particle into the 1064 nm trap of the main tweezer is achieved by slowly decreasing the 976 nm laser power while increasing that of the 1064 nm laser. We then turn on the 976 nm trap of the main tweezer. By using the electrode shown if Fig. 1(c) of the main text, we apply an electrostatic force on the particle to push it and make it jump to the 976 nm trap. To load the second particle, we bring the collector back in the first chamber, we capture another particle and move it to the 1064 nm focal point of the main tweezer by first maximizing the 1064 nm laser power transmitted through the collector, and then readjusting its position to account for chromatic aberration. By decreasing the power in the collector trap, the particle is trapped at the 1064 nm focal point. At the end of the procedure, both particles are trapped in the same tweezer, but at different capture points.

During the whole procedure the alignment between the two optical systems is never optimized with both 976 nm lasers turned on simultaneously, thus avoiding a configuration that could damage the lasers.

Calculating the exact focal point requires accurate knowledge of the lens shapes, but the shift in its position due to chromatic aberration can be estimated using geometrical optics for thin lenses made of the same DZK-3 glass, as follows. The light exiting the fiber is collimated and re-focused by a pair of lenses with focal lengths $f_1$ and $f_2$ respectively. We assume thin lenses at close distance, and a dependence of the focal length $f$ on the refractive index $n$ as $f \propto 1/(n-1)$, such that

$$\frac{\delta f}{f} = -\frac{\delta n}{n-1} \, . \tag{Eq.S1}$$

If $p$ and $q$ are the distances from a lens of the object and the image respectively, the thin lens formula is

$$1/p + 1/q = 1/f \tag{Eq.S2}$$

We considered the system optimized for a wavelength $a$, so that $p_1^a = f_1$, $q_1^a = p_2^a = \infty$, and $q_2^a = f_2$. For a second wavelength $b$, we have $p_1^b = p_1^a$ (this is the physical distance between fiber and lens) and $f_i^b = f_i + \delta f_i$, so that, e.g., for the first lens (Eq.S2) becomes $1/f_1 + 1/q_1^b = 1/(f_1 + \delta f_1)$. After some algebra, keeping the equations at the first order in $\delta f$, we get

$$\delta q_2 = -\frac{\delta n}{n-1}\left(f_2 + \frac{f_2^2}{f_1}\right). \tag{Eq.S3}$$

In our system the lenses are made of D-ZK3 glass, having $n_{1064} = 1.5771$ and $n_{976} = 1.5785$. If we take 1064 nm as wavelength $a$, we have therefore $\delta n = 0.0014$ and $\delta n/(n-1) = 2.4 \cdot 10^{-3}$. The first tweezer has $f_1 = 18.4$mm and $f_2 = 3.1$mm, therefore (Eq.S3) gives $\delta q_2 = -8.8\,\mu$m.

By measuring the displacement of the tweezer optics between the positions that optimize the coupling of the light to the collector for the two wavelengths, we deduced in Ref. [1] a $9 \pm 1\,\mu$m shift of the focus, in agreement with the calculation of the chromatic aberration.

## Supplementary 2 : Optical binding interaction

In this section we evaluate the optical binding between particles in our experiment. The optical binding arises from the interaction of one particle with the dipolar radiation (Rayleigh scattered tweezer light) of the other and vice versa. The field-dipole interaction results in a force that each particle applies on the other. A treatment of the optical binding can be found in the supplementary materials of [2], where the authors use a Green tensor to calculate the total field at the positions of the particles for two parallel tweezers. In our experiment the two particles are both positioned along the optical axis of the same tweezer, and the whole interaction can be derived using a Gaussian approximation for focused beams. The optical potential for a spherical dielectric particle is

$$U = -\alpha \frac{I}{2\epsilon_0 c} \tag{Eq.S4}$$

where $I$ is the radiation intensity and $\alpha = V\epsilon_0\chi$ is the polarizability, $V$ the particle volume and $\chi = 3\dfrac{\epsilon-1}{\epsilon+3}$ the optical susceptibility. Considering a laser field $E_0$ linearly polarized along the axis $\hat{y}$ at the nanosphere position, the scattered field in the $\hat{x}\hat{z}$ plane at a distance $r$ from the particle reads

$$\mathbf{E}_{sc} = -\frac{1}{4\pi r}k^2\frac{\alpha}{\epsilon_0}e^{ikr}E_0\,\hat{y}. \tag{Eq.S5}$$

In our experimental configuration each particle interacts with two different laser fields. Consequently, we need to consider the force induced by the scattering of each laser light by each particle, as done in the following sections.

*Force on particle P2 due to the 976nm laser*

We first consider the optical potential on the particle P2 due to the first laser (at 976nm). The field at the position $\mathbf{r}_2$ of particle P2 is the coherent sum of the tweezer light and the light scattered by particle P1: $E_0(\mathbf{r}_2) = E_{tw}(\mathbf{r}_2) + E_{sc}(\mathbf{r}_2)$ where the last term is produced by $E_{tw}(\mathbf{r}_1)$ scattered by particle P1. In $|E_0(\mathbf{r}_2)|^2$, the relevant term is $2\text{Re}\left[E_{tw}^*E_{sc}\right]$. We can calculate the absolute values of the fields at the respective particle equilibrium positions, distant $d$, and keep the dependence on the particles coordinates just in the phases, such as $E_{tw}(\mathbf{r}_1) \simeq |E_{tw}(0,0,0)|e^{i\phi_{tw}(\mathbf{r}_1)}$, $E_{tw}(\mathbf{r}_2) \simeq |E_{tw}(0,0,d)|e^{i\phi_{tw}(\mathbf{r}_2)}$ and, in the scattered field, $1/r \simeq 1/d$. In the phase term $ikr$, we can approximate

$$r \simeq d + z_2 - z_1 + \frac{(y_2-y_1)^2}{2d} + \frac{(x_2-x_1)^2}{2d}. \tag{Eq.S6}$$

For the phase of the tweezer light, we use the Gaussian field approximation

$$\phi_{tw}(\mathbf{r_1}) \simeq kz_1 - \arctan\frac{z_1}{z_R}$$

$$\phi_{tw}(\mathbf{r_2}) \simeq k(d + z_2) - \arctan\frac{d + z_2}{z_R} + \frac{k}{2(d + z_R^2/d)}(x_2^2 + y_2^2). \quad \text{(Eq.S7)}$$

Finally, we have

$$U = -\frac{\alpha_2}{2\epsilon_0 c}\left(-2\frac{\alpha_1 k^2}{4\pi\epsilon_0 d}\right)\sqrt{I_{tw}(0)I_{tw}(d)}\cos\Phi \quad \text{(Eq.S8)}$$

where

$$\Phi = \left(kr + \phi_{tw}(\mathbf{r_1})\right) - \phi_{tw}(\mathbf{r_2})$$

$$\simeq \arctan\frac{d + z_2}{z_R} - \arctan\frac{z_1}{z_R} + \frac{k[(x_2 - x_1)^2 + (y_2 - y_1)^2]}{2d} - \frac{k(x_2^2 + y_2^2)}{2(d + z_R^2/d)} \quad \text{(Eq.S9)}$$

In the above expression, the only term that is not $\ll 1$ is the first one, such that, writing $\Phi = \Phi_0 + \delta\Phi$ with $\Phi_0 = \arctan d/z_R$, we can approximate $\cos\Phi \simeq \cos\Phi_0 - \sin\Phi_0 \, \delta\Phi = \frac{1}{\sqrt{1+(d/z_R)^2}} - \frac{1}{\sqrt{1+(z_R/d)^2}}\delta\Phi$.

The force acting of particle P2 is calculated as the gradient of $U$ with respect to $\mathbf{r}_2$ (namely, $F = -\nabla U$), and we are actually interested in the terms of $F_z$ proportional to either $z_1$ or $z_2$ (and the same for $x$ and $y$). Therefore, in $U$ we just keep the terms proportional to $r_{2,i}^2$ or $r_{2,i}r_{1,i}$ with $i = x, y, z$:

$$\cos\Phi \to \frac{1}{\sqrt{1 + (z_R/d)^2}}\left(\frac{k}{d}(x_1 x_2 + y_1 y_2) - \frac{k}{2d}\frac{z_R^2}{d^2 + z_R^2}(x_2^2 + y_2^2)\right). \quad \text{(Eq.S10)}$$

Finally, we obtain a force

$$F^x = k_{1\to 2}\, x_1 - \frac{z_R^2}{d^2 + z_R^2}k_{1\to 2}\, x_2 \quad \text{(Eq.S11)}$$

with

$$k_{1\to 2} = -\frac{\alpha_1\alpha_2 k^3}{4\pi\epsilon_0^2\, c\, d^2}\sqrt{I_{tw}(0)I_{tw}(d)}\,\frac{1}{\sqrt{1 + (z_R/d)^2}} \quad \text{(Eq.S12)}$$

and similar expressions for $F^y$, while at this order there is no optical binding for the $z$ direction.

*Force on particle P2 due to the 1064nm laser*

The phase of the tweezer field is now

$$\phi_{tw}(\mathbf{r_1}) \simeq k(-d + z_1) - \arctan\frac{(-d + z_1)}{z_R} - \frac{k}{2(d + z_R^2/d)}(x_1^2 + y_1^2)$$

$$\phi_{tw}(\mathbf{r_2}) \simeq kz_2 - \arctan\frac{z_2}{z_R} \quad \text{(Eq.S13)}$$

therefore

$$\Phi \simeq \arctan\frac{d - z_1}{z_R} + \arctan\frac{z_2}{z_R} + \frac{k[(x_2 - x_1)^2 + (y_2 - y_1)^2]}{2d} - \frac{k(x_1^2 + y_1^2)}{2(d + z_R^2/d)} \quad \text{(Eq.S14)}$$

and
$$\cos \Phi \to \frac{1}{\sqrt{1+(z_R/d)^2}} \left( \frac{k}{d}(x_1 x_2 + y_1 y_2) - \frac{k}{2d}(x_2^2 + y_2^2) \right) \tag{Eq.S15}$$

giving a force
$$F^x = k_{1\to 2}(x_1 - x_2) \tag{Eq.S16}$$

with the same $k_{1\to 2}$ given in (Eq.S12) (of course, with different laser parameters), and similar expressions for $F^y$.

### Force on particle P1 due to the 976nm laser

The optical potential is the same as in (Eq.S8), with $\Phi$ given by
$$\Phi = \left( kr + \phi_{tw}(\mathbf{r}_2) \right) - \phi_{tw}(\mathbf{r}_1) \tag{Eq.S17}$$

where the phase of the tweezer field is given in (Eq.S7), therefore
$$\begin{aligned}\Phi \simeq{} & 2k(d + z_2 - z_1) - \arctan\frac{d+z_2}{z_R} + \arctan\frac{z_1}{z_R} \\ & + \frac{k}{2d}[(x_2 - x_1)^2 + (y_2 - y_1)^2] + \frac{k}{2(d + z_R^2/d)}(x_2^2 + y_2^2).\end{aligned} \tag{Eq.S18}$$

The term that is not $\ll 1$ is now
$$\Phi_0 = 2kd - \arctan\frac{d}{z_R} \tag{Eq.S19}$$

and $\cos \Phi$ is still approximated as $\cos \Phi \simeq \cos \Phi_0 - \sin \Phi_0\, \delta\Phi$.

We have now to calculate the gradient of $-U$ with respect to $\mathbf{r}_1$, therefore we keep the terms proportional to $r_{1,i}^2$ or $r_{1,i} r_{2,i}$:
$$\cos \Phi \to \sin \Phi_0 \left( \frac{k}{d}(x_1 x_2 + y_1 y_2) - \frac{k}{2d}(x_1^2 + y_1^2) \right) \tag{Eq.S20}$$

giving a force
$$F^x = k_{2\to 1}(x_2 - x_1) \tag{Eq.S21}$$

with
$$k_{2\to 1} = -\frac{\alpha_1 \alpha_2 k^3}{4\pi\epsilon_0^2\, c\, d^2} \sqrt{I_{tw}(0) I_{tw}(d)}\, \sin \Phi_0 \tag{Eq.S22}$$

and similar expressions for $F^y$.

### Force on particle P1 due to the 1064nm laser

The phase $\Phi$ is the one in (Eq.S17) with the tweezer field phase of (Eq.S13), therefore
$$\begin{aligned}\Phi \simeq{} & 2k(d + z_2 - z_1) - \arctan\frac{d-z_1}{z_R} - \arctan\frac{z_2}{z_R} \\ & + \frac{k}{2d}[(x_2 - x_1)^2 + (y_2 - y_1)^2] + \frac{k}{2(d + z_R^2/d)}(x_1^2 + y_1^2)\end{aligned} \tag{Eq.S23}$$

and
$$\cos \Phi \to \sin \Phi_0 \left( \frac{k}{d}(x_1 x_2 + y_1 y_2) - \frac{k}{2d}\frac{2+(z_R/d)^2}{1+(z_R/d)^2}(x_1^2 + y_1^2) \right). \tag{Eq.S24}$$

The force is
$$F^x = k_{2\to 1}\left( x_2 - \frac{2+(z_R/d)^2}{1+(z_R/d)^2} x_1 \right) \tag{Eq.S25}$$

with the same $k_{2\to 1}$ given in (Eq.S22), and similar expressions for $F^y$.

*Conclusions*

We can define the reciprocal and anti-reciprocal interaction coefficients [2] respectively as

$$k_1 = 0.5(k_{1\to 2} + k_{2\to 1}) \tag{Eq.S26}$$
$$k_2 = 0.5(k_{1\to 2} - k_{2\to 1}) \tag{Eq.S27}$$

so that the evolution equations can be written as

$$m\ddot{x}_1 + m\gamma\dot{x}_1 = -m\Omega_1^2 x_1 + (k_1 + k_2)x_2 \tag{Eq.S28}$$
$$m\ddot{x}_2 + m\gamma\dot{x}_2 = -m\Omega_2^2 x_2 + (k_1 - k_2)x_1 \tag{Eq.S29}$$

where the eigenfrequencies $\Omega_{1,2}$ are modified with respect to the original ones according to the appropriate terms in the expression of $\sum F^x$.

As in [2], at this order there is no optical binding along the direction connecting the two particles (the tweezer axis $z$ for our work, the direction $x$ for [2]). This is due to the expansion of $r$, lacking of the second order term in $z$. A binding force along $z$ can appear if we expand the Gouy phase (that we keep here at the 0 order).

With respect to [2], our coupling is lower by a factor of $\sqrt{I_{tw}(d)/I_{tw}(0)}$ because in [2] the tweezer field is close to the focus for both particles, and by a further factor $1/kd$ because in [2] the tweezer phase is linear with the relevant variable $z$.

The quantitative evaluation of all equivalent optical binding spring constant is summarized in the notebook reported below. It contains the geometrical properties of the laser beams, assumed as Gaussian. The values of the waists and Rayleigh ranges are calculated with the nominal lenses and fiber geometry. We emphasize that, close to a focal point, the Gaussian approximation does not hold. As a consequence, the nominal values of the waist and the Rayleigh range for one laser are only suited to calculate the field at the position of the particle trapped by the other laser. For the field felt and scattered by the nanosphere trapped in the focus we use instead effective parameters deduced from the measured trapping frequency. We find that the optical binding interaction is more than two orders of magnitude lower than the Coulomb interaction, and we neglect it in our analysis of the experimental results.

# Evaluation of the optical binding;

## Particles parameters;

```
In[•]:= c = 299 792 458; (* Light velocity *)
       ϵ0 = 8.854187812 × 10⁻¹²; (* Vacuum permittivity *)
       rho = 1850; (* Particles density *)
       r = 125 × 10⁻⁹ / 2; (* Particles radius *)
       V = 4 / 3 π r³; (* Particles volume *)
       m = rho V; (* Particles mass *)
       α = α1 = α2 = V ϵ0 χ; (* Particles polarizability *)
       ϵ = 2.25; (* Particles relative permittivity *)
       χ = 3 (ϵ - 1)/(ϵ + 3); (* Parlticles optical susceptibility *)
       r12 = 8.8 × 10⁻⁶ ; (* Distance separating the two particles *)

       ωZ₁₀₆₄ = 2 π 20.2 × 10³ ;
       (* Bare Z frequency of the particle trapped in the 1064nm laser *)
       ωY₁₀₆₄ = 2 π 108.5 × 10³ ;
       (* Bare Y frequency of the particle trapped in the 1064nm laser *)
       ωX₁₀₆₄ = 2 π 114.8 × 10³;
       (* Bare X frequency of the particle trapped in the 1064nm laser *)
       ωZ₉₇₆ = 2 π 26.6 × 10³ ;
       (* Bare max. frequency of the particle trap in the 976nm laser *)
       ωY₉₇₆ = 2 π 147.8 × 10³;
       (* Bare Y frequency of the particle trapped in the 1064nm laser *)
       ωX₉₇₆ = 2 π 130 × 10³; (* Bare X frequency of the particle trapped in the 1064nm laser *)
```

## Lasers parameters;

```
In[•]:= λ₁₀₆₄ = 1064 × 10⁻⁹; (* Wavlength of the 1064 nm laser *)
       k₁₀₆₄ = 2π/λ₁₀₆₄; (* Corresponding wave number *)
       λ₉₇₆ = 976 × 10⁻⁹; (* Wavelength of the 976 nm laser *)
       k₉₇₆ = 2π/λ₉₇₆; (* Corresponding wave number *)

       zR₁₀₆₄ = 1.00 × 10⁻⁶; (* Nominal Rayleigh range of the 1064nm focused laser beam *)
       zR₉₇₆ = 0.92 × 10⁻⁶; (* Nominal Rayleigh range of the 976nm focused laser beam *)
       w0₁₀₆₄ = 0.58 × 10⁻⁶; (* Nominal waist of the 1064nm focused laser beam *)
       w0₉₇₆ = 0.53 × 10⁻⁶; (* Nominal waist of the 976nm focused laser beam *)
       P0₁₀₆₄ = 250 × 10⁻³; (* Optical power in the 1064nm trap *)
       ϕ0₉₇₆ = 2 k₉₇₆ r12 - ArcTan[r12/zR₉₇₆];
       (* Phase of the 976nm laser at a distance r12 from the focus *)
       ϕ0₁₀₆₄ = 2 k₁₀₆₄ r12 - ArcTan[r12/zR₁₀₆₄];
       (* Phase of the 1064nm laser at a distance r12 from the focus *)
       P0₉₇₆ = P0₁₀₆₄ √(ωZ₉₇₆/ωZ₁₀₆₄) ; (* Max. optical power in the 976nm
        trap. Evaluated from the measured frequency ratio between P1 and P2 *)
```

$$I0_{1064} = \frac{2\,P0_{1064}}{\pi\,w0_{1064}^2};\ (*\ \text{Intensity at the focus determined using Gaussian approximation for the 1064nm laser }*)$$

$$I0_{976} = \frac{2\,P0_{976}}{\pi\,w0_{976}^2};\ (*\ \text{Intensity at the focus determined using Gaussian approximation for the 976nm laser }*)$$

$$WY0_{1064} = \frac{1}{\omega Y_{1064}}\sqrt{\frac{I0_{1064}\,\alpha}{m\,\epsilon 0\,c}};\ (*\ \text{Effective beam waist along Y, at the 1064nm focus, obtained from the measured oscillators frequency }*)$$

$$WX0_{1064} = \frac{1}{\omega X_{1064}}\sqrt{\frac{I0_{1064}\,\alpha}{m\,\epsilon 0\,c}};\ (*\ \text{Effective beam waist along X, at the 1064nm focus, obtained from the measured oscillators frequency }*)$$

$$WY0_{976} = \frac{1}{\omega Y_{976}}\sqrt{\frac{I0_{976}\,\alpha}{m\,\epsilon 0\,c}};\ (*\ \text{Effective beam waist along Y, at the 976nm focus, obtained from the measured oscillators frequency }*)$$

$$WX0_{976} = \frac{1}{\omega X_{976}}\sqrt{\frac{I0_{976}\,\alpha}{m\,\epsilon 0\,c}};\ (*\ \text{Effective beam waist along X, at the 976nm focus, obtained from the measured oscillator's frequency }*)$$

$$W0_{1064} = \frac{WX0_{1064} + WY0_{1064}}{2};\ (*\ \text{Effective average waist at the 1064nm focus, used to calculate the intensity at the focal point }*)$$

$$W0_{976} = \frac{WX0_{976} + WY0_{976}}{2};\ (*\ \text{Effective average waist at the 976nm focus, used to calculate the intensity at the focal point }*)$$

$$i0_{1064} = \frac{2\,P0_{1064}}{\pi\,W0_{1064}^2};\ (*\ \text{Intensity at the focus determined using the effective waist of the 1064nm laser beam }*)$$

$$i0_{976} = \frac{2\,P0_{976}}{\pi\,W0_{976}^2};\ (*\ \text{Intensity at the focus determined using the effective waist of the 976nm laser beam }*)$$

$$Itw_{1064}[z\_] := I0_{1064}\left(\frac{w0_{1064}}{w_{1064}[z]}\right)^2;\ (*\ \text{Nominal 1064nm laser intensity at the beam center as a function of the distance to the focus }*)$$

$$w_{1064}[z\_] := w0_{1064}\sqrt{1+\left(\frac{z}{zR_{1064}}\right)^2};\ (*\ \text{Nominal 1064nm laser beam waist as a function of the distance to the focus (Gaussian approximation) }*)$$

$$Itw_{976}[z\_] := I0_{976}\left(\frac{w0_{976}}{w_{976}[z]}\right)^2;\ (*\ \text{Nominal 976nm laser intensity at the beam center as a function of the distance to the focus }*)$$

$$w_{976}[z\_] := w0_{976}\sqrt{1+\left(\frac{z}{zR_{976}}\right)^2};\ (*\ \text{Nominal 976nm laser beam waist as a function of the distance to the focus (Gaussian approximation)}*)$$

## Expression of the equivalent spring constants of the optical binding interaction;

$$\text{In[}\circ\text{]:= } k12_{976} = -\frac{\alpha1\,\alpha2\,k_{976}^3}{4\pi\,\epsilon0^2\,c\,r12^2}\sqrt{i0_{976}\,\text{Itw}_{976}[r12]}\,\frac{1}{\sqrt{1+(zR_{976}/r12)^2}};$$

(* Equivalent spring constant due to the 976nm scattered light on particle P2 *)

$$k12_{1064} = -\frac{\alpha1\,\alpha2\,k_{1064}^3}{4\pi\,\epsilon0^2\,c\,r12^2}\sqrt{i0_{1064}\,\text{Itw}_{1064}[r12]}\,\frac{1}{\sqrt{1+(zR_{1064}/r12)^2}};$$

(* Equivalent spring constant due to the 1064nm scattered light on particle P2 *)

$$k21_{976} = -\frac{\alpha1\,\alpha2\,k_{976}^3}{4\pi\,\epsilon0^2\,c\,r12^2}\sqrt{i0_{976}\,\text{Itw}_{976}[r12]}\,\text{Sin}[\phi0_{976}];$$

(* Equivalent spring constant due to the 976nm scattered light on particle P1 *)

$$k21_{1064} = -\frac{\alpha1\,\alpha2\,k_{1064}^3}{4\pi\,\epsilon0^2\,c\,r12^2}\sqrt{i0_{1064}\,\text{Itw}_{1064}[r12]}\,\text{Sin}[\phi0_{1064}];$$

(* Equivalent spring constant due to the 1064nm scattered light on particle P2 *)

## Print the equivalent spring constant for the optical binding interaction;

In[◦]:= k12$_{976}$
k21$_{976}$
k12$_{1064}$
k21$_{1064}$

Out[◦]= $-1.65169 \times 10^{-11}$

Out[◦]= $1.58144 \times 10^{-11}$

Out[◦]= $-1.04119 \times 10^{-11}$

Out[◦]= $-9.75839 \times 10^{-12}$

## Coulomb coupling;

In[◦]:= qe = $1.602176634 \times 10^{-19}$; (* Elementary charge *)
q1 = 67 qe; (* Charge of particle P1 *)
q2 = 90 qe; (* Charge of particle P2 *)
$kz_{\text{Coul}} = 2\,\frac{q1\,q2}{4\pi\,\epsilon0\,r12^3}$ (* Equivalent spring constant due to the Coulomb coupling *)

Out[◦]= $4.08283 \times 10^{-9}$

## Comparison between the X–Y optical binding and the Z Coulomb coupling;

*In[ ]:=* **k12$_{976}$ / kz$_{Coul}$**
**k21$_{976}$ / kz$_{Coul}$**
**k12$_{1064}$ / kz$_{Coul}$**
**k21$_{1064}$ / kz$_{Coul}$**

*Out[ ]=* $-0.00404545$

*Out[ ]=* $0.0038734$

*Out[ ]=* $-0.00255016$

*Out[ ]=* $-0.0023901$

## Supplementary 3 : Calibration of the charge numbers

To estimate the number of charges on a nanoparticle, we used a similar method as reported in Ref. [3]. We implemented in the loading chamber an antenna extending close to the tweezer optics, onto which we apply a DC voltage of ≈ 500 V. At a pressure of 1.5 mbar, this voltage ionizes the nitrogen atmosphere in the chamber, leading to an increase in positive or negative charges on the surface of the nanosphere. During the application of the electrical discharge, we drive the particle motion by using the ring electrode to apply an oscillating field at the Z motion frequency. The photodetection signal of the back-reflected light is then demodulated at the driving frequency, obtaining a signal proportional to the particle charge. We can thus monitor the charge jumps in real time, as shown in Fig. S1 (a). In this figure the data are sorted with respect to the demodulated amplitude in order to emphasize the jumps. On the figure we can see some individual steps of identical height, which we interpret as single charge jumps. Since a neutral particle gives a null signal, we can use them to directly calibrate the charge number.

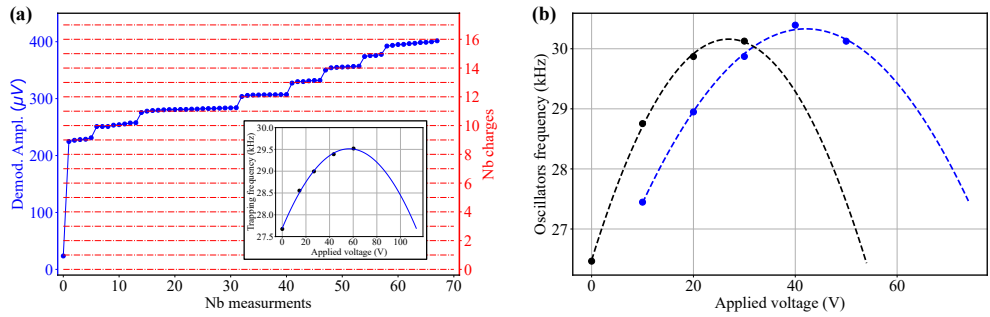

Fig. S1. **Calibration of the number of charges:** (a) Demodulated signal of the particle motion under coherent drive (10 V peak-to-peak), while an electric discharge is turned on. We observe discrete steps of repeatable size, corresponding to single charge jumps. The inset shows a measurement of the trapping frequency as a function of the applied static voltage (i.e., as a function of the position in the trapping potential), with a fit to a quadratic behaviour. The aperture of the parabola is proportional to the square root of the electrical charge. (b) Measurement of the Z-motion frequencies as a function of the applied voltage for the two particles used in the experiment described in the main text. Comparing the parabola aperture with that of the reference particle (see Fig. S1(a)), for which the number of charges is known, we can infer the charge numbers for the two nanospheres of interest.

We note that, beyond 15 charges, the size of the jumps starts to decrease. We explain this phenomenon by the non-harmonicity of the trapping potential which cannot be considered as quadratic far from the center. As a consequence, the oscillator frequency is position-dependant, with a maximum at the trap center. For a large number of charges the motion amplitude increases and the particle starts to explore a wider range of the trap, spending more time in regions far from the center, which lowers its resonance frequency. As a consequence, the driving force is no more at the maximum of the oscillator susceptibility, the motion amplitude is reduced, and the jump size in the demodulated signal decreases.

In order to extend the calibration protocol to other nanospheres without always repeating the entire procedure described above, we exploit the anharmonicity of the trap as follows. We apply a static electric field on the charged particle, which modifies its equilibrium position and therefore shifts its resonance frequency. The latter, as a function of the static voltage applied to the ring electrode, is shown in the inset of figure S1(a), where a clear quadratic dependence is visible. Assuming, to the lowest order, a linear dependence of the position on the applied electric force and a quadratic dependence of the resonance frequency around its maximum, we can write the resonance frequency as $f_{max}[1 - A(V - V_0)^2]$ where $V$ is the applied voltage. $f_{max}$ and the offset $V_0$ depend on the particle properties, and $V_0$ can be attributed to stray fields and to radiation pressure. The quadratic coefficient $A$, is proportional to the square of the charge and depends on the system geometry and the optical potential. We note that, for a sample of 20 particles from the same batch, we observed a very reproducible value of $f_{max}$ with a relative standard deviation of less than one percent. This confirms the stability of our trapping system in terms of laser power and geometry.

For the reference particle we arrived to a charge of $|q| = 30 q_e$ (following the steps in the demodulated amplitude), then we turned off the discharge and measured the resonance frequency as a function of the static voltage, obtaining a quadratic coefficient of $(1.96 \pm 0.11) \cdot 10^{-5}$ V$^{-2}$

We performed the latter measurement for the two particles used in the Coulomb coupling experiment, as shown in figure S1 (b). Our fibered setup ensures the reproducibility of the trap shape, so that we can deduce the particle charges by comparing the quadratic coefficients. We found $(9.3 \pm 1.5) \cdot 10^{-5}$ V$^{-2}$ for P1 and $(1.7 \pm 0.4) \cdot 10^{-4}$ V$^{-2}$ for P2. Taking into account the uncertainties we deduce $|q_1| = (67 \pm 7) \, q_e$ and $|q_2| = (90 \pm 13) \, q_e$.

By monitoring the response to the modulated electric field, we deduce that the charge number remains stable after the discharge is turned off and during the translation of the tweezer to the science chamber.

## Supplementary 4 : Spectrum of two coupled particles

The Newton's equations in the time domain for two reciprocally coupled harmonic oscillators driven by thermal force $F_{1,2}^{th}$ are

$$
\begin{aligned}
m_1 \ddot{z}_1 &= -\left(m_1 \Omega_{Z1}^2 + k\right) z_1 + k z_2 - m_1 \gamma_1 \dot{z}_1 + F_1^{th} \\
m_2 \ddot{z}_2 &= -\left(m_2 \Omega_{Z2}^2 + k\right) z_2 + k z_1 - m_2 \gamma_2 \dot{z}_2 + F_2^{th}
\end{aligned}
\quad \text{(Eq.S30)}
$$

We can express them in the Fourier domain as

$$
\begin{aligned}
\frac{\tilde{F}_1^{th}}{m_1} &= \left(\Omega_1^2 - \omega^2 - i\omega\gamma_1\right) \tilde{z}_1 - \frac{k}{m_1} \tilde{z}_2 \\
\frac{\tilde{F}_2^{th}}{m_2} &= \left(\Omega_2^2 - \omega^2 - i\omega\gamma_2\right) \tilde{z}_2 - \frac{k}{m_2} \tilde{z}_1
\end{aligned}
\quad \text{(Eq.S31)}
$$

where $\tilde{x}$ represents the Fourier transformed of a time-limited sample of the stochastic variable $x$, and $\Omega_i^2 = \Omega_{Zi}^2 + k/m_i$ with $i = 1, 2$. These equations can be written as

$$
\begin{aligned}
s_1 \tilde{z}_1 - c\mu \tilde{z}_2 &= \frac{\tilde{F}_1^{th}}{m_1} \\
s_2 \tilde{z}_2 - \frac{c}{\mu} \tilde{z}_1 &= \frac{\tilde{F}_2^{th}}{m_2}
\end{aligned}
\quad \text{(Eq.S32)}
$$

by defining $\mu = \sqrt{m_2/m_1}$, $c = k/\sqrt{m_1 m_2}$ and $s_i = -\omega^2 - i\omega\gamma_i + \Omega_i^2$ with $\Omega_1^2 = \Omega_{Z1}^2 + c\mu$ and $\Omega_2^2 = \Omega_{Z2}^2 + \frac{c}{\mu}$.

Since the photo-detector produces a signal $z_{out}(t) = A z_1(t) + B z_2(t)$, we can diagonalize the system to get the expressions of $\tilde{z}_{1,2}$, and derive the power spectrum $S_{out}$ of $z_{out}$. Assuming that

the noise forces $F_i^{th}$ are uncorrelated and have spectral densities $S_i$, we obtain

$$S_{out} = \frac{\left|As_2 + B\frac{c}{\mu}\right|^2 \bar{S}_1 + |Bs_1 + Ac\mu|^2 \bar{S}_2}{|s_1 s_2 - c^2|^2} \quad \text{(Eq.S33)}$$

where $\bar{S}_i = S_i/m_i^2$ are the acceleration noise power spectra. Eq. (Eq.S33) cannot readily be used to fit the experimental spectra. It is indeed under-determined since multiplying $A$ and $B$ by a constant and dividing $\bar{S}_1$ and $\bar{S}_2$ by its square root leaves $S_{out}$ unchanged. We can then re-define $A\sqrt{\bar{S}_1} \to A$ and $B\sqrt{\bar{S}_1} \to B$ and re-write Eq. (Eq.S33) as

$$S_{out} = \frac{\left|As_2 + B\frac{c}{\mu}\right|^2 + |Bs_1 + Ac\mu|^2 \bar{S}_2/\bar{S}_1}{|s_1 s_2 - c^2|^2} \,. \quad \text{(Eq.S34)}$$

Assuming the driving noise to be thermal and an equal temperature for the two particles we have, according to the fluctuation-dissipation theorem, $\frac{\bar{S}_2}{\bar{S}_1} = \frac{\gamma_2}{\gamma_1 \mu^2}$. We then obtain the final equation

$$S_{out} = \frac{\left|As_2 + B\frac{c}{\mu}\right|^2 + |Bs_1 + Ac\mu|^2 \frac{\gamma_2}{\gamma_1 \mu^2}}{|s_1 s_2 - c^2|^2} \quad \text{(Eq.S35)}$$

that is reported in the main text.

**References**

1. M. Calamai, A. Ranfagni, and F. Marin, "Transfer of a levitating nanoparticle between optical tweezers," AIP Adv. **11**, 025246 (2021).
2. J. Rieser, M. A. Ciampini, H. Rudolph, *et al.*, "Tunable light-induced dipole-dipole interaction between optically levitated nanoparticles," Science **377**, 987–990 (2022).
3. M. Frimmer, K. Luszcz, S. Ferreiro, *et al.*, "Controlling the net charge on a nanoparticle optically levitated in vacuum," Phys. Rev. A **95**, 061801 (2017).